\documentclass[10pt]{IEEEtran}
\usepackage[utf8]{inputenc}
\usepackage{epsfig}
\usepackage{subfigure}
\usepackage{hyperref}
\usepackage{graphicx}
  \DeclareGraphicsExtensions{.eps}
\usepackage{amssymb}

\newenvironment{mylist}[1][$\bullet$]{\begin{list}{#1}{\leftmargin \parindent \itemindent 0mm \labelwidth \parindent}}{\end{list}}

\title{Analysis of Quasi-Cyclic LDPC codes under ML decoding over the erasure channel}

\author{
 Mathieu Cunche$^{\ast}$,  Valentin Savin$^{\natural}$, Vincent Roca$^{\ast}$ \\
  \begin{small} $^{\ast}$INRIA Rh\^one-Alpes, Grenoble, France, $^{\natural}$CEA-LETI MINATEC, Grenoble, France                                                                                                                                \end{small}
\thanks{This work was supported by the French ANR grant No 2006 TCOM 019 (CAPRI-FEC project).}}

\begin{document}

\makeatother
\maketitle

\begin{abstract}

In this paper, we show that Quasi-Cyclic LDPC codes can efficiently accommodate the hybrid iterative/ML decoding over
the binary erasure channel. We demonstrate that the quasi-cyclic structure of the parity-check matrix can be
advantageously used in order to significantly reduce the complexity of the ML decoding. This is achieved by a simple
row/column permutation that transforms a QC matrix into a pseudo-band form. Based on this approach, we propose a class
of QC-LDPC codes with almost ideal error correction performance under the ML decoding, while the required number of
row/symbol operations scales as $k\sqrt{k}$, where $k$ is the number of source symbols.

\end{abstract}

\section{Context and related works}
\label{sec:intro}
In modern communication systems, data is often transmitted as independent packets. These packets can
be subject to losses (erasures) caused by bad channel conditions, intermittent connectivity, congested routers, or
failures. If solutions based on the retransmission of lost packets are possible (ARQ, Automatic Repeat Requests), they
are not always suitable (e.g. broadcasting), nor possible (no return link, e.g. satellite communications). In such
cases Forward Error Correction (FEC) schemes represent  the foremost alternative. These schemes rely on erasure codes
operating either at the transport or the application layer of the communication system, which are able to recover lost
data thanks to the transmission of redundant (repair) packets.

In the family of error-correcting codes, a prominent role is played by Low-Density Parity-Check (LDPC) codes. They
feature a linear complexity iterative (IT) decoding, and can be optimized for a broad class of channels, with
asymptotically performance close to the theoretical Shannon limit. Although iterative and maximum likelihood (ML) are
equivalent for cycle-free codes, for a given finite code (with cycles) the gap between their performance can be
significant. Hence, ML decoding has been recently considered in order to improve the correction capacity of LDPC codes
over the binary erasure channel (BEC) for short to moderate code-length. This comes at a cost in the decoding
complexity; however, efficient ML decoding algorithms with reduced complexity have been proposed over the last few
years \cite{Burshtein04}.

Before discussing the complexity of the ML decoding, let us first consider the complexity of the encoding process.
Encoding a systematic LDPC code is equivalent to solving a linear system $H_pP = H_sS$, where $H = (H_s, H_p)$ is the
parity-check matrix of the code, and $S$ and $P$ denote respectively the sequences of source and parity bits. This can
be done by Gaussian elimination (GE), whose complexity\footnote{We consider here the complexity of the GE, and not of
the encoding process itself. Clearly GE is performed only once, and can be done ``offline'', hence its complexity is
irrelevant for the encoding process itself, but it is relevant in the perspective of the subsequent discussion about ML
decoding complexity.}, expressed as the number of {\em row operations}\footnote{Each row operation requires $k$ bit 
operations (corresponding to the $k$ entries of the row), and one operation on the right-hand side of the system. For
Application Layer (AL)-FEC, the right-hand side is not a bit, but an entire packet, also called symbol. Thus, a row
operation will be also referred to as {\em symbol operation}}, is expected to scale as $k^2$, where $k$ denotes the
number of source bits. However, it has been shown in \cite{RichUrb01} that the GE can take advantage of the sparseness
of the parity check matrix, and it can be efficiently performed in ${\cal O}((gk)^2)$ row/symbol operations, where $g$
is called the {\em gap of the code}. Roughly speaking, the idea behind is that if a fraction $g$ of parity bits are
resolved, remaining parity bits can be recovered by performing an iterative erasure decoding.

Similar considerations apply to the ML decoding over the BEC, which consists of solving the linear {\em residual
system} $H_e X_e = H_r X_r$, where $X_r$ and $X_e$ denote the vectors of received and of erased bits, respectively, and
$H_r$, $H_e$ are the corresponding submatrices of $H$. Using a GE algorithm that takes advantage of the sparseness of
this system \cite{LaMacchia91}, \cite{Burshtein04}, the decoding complexity scales, {\em in average}, as $(\varepsilon
k)^2$ row/symbol operations, where $\varepsilon$ is the average reception overhead necessary to successfully complete
the iterative decoding. However, the decoding complexity is still quadratic in $k$. As the code length tends to
infinity, $\varepsilon$ tends to a positive threshold value, but even if this asymptotic threshold is close to $0$,
$\varepsilon$ still can be relatively large for finite codes. Besides, typically, there is a tradeoff between the
performance of the IT decoding that of the ML decoding. Consequently, improvement of the ML decoding performance comes
at the price of some degradation of the IT performance, which results in an increased average overhead
$\varepsilon_{IT}$ \cite{paolini-2008b}.
For instance, for regular repeat-accumulate (RRA) codes, it has been shown in \cite{cunche08hybride} that increasing the degree of source 
bit-nodes results in an improvement of the ML performance, but induces a degradation of the  IT performance. Hybrid
IT-ML decoding algorithms have also been considered in \cite{paolini-2008a}.

Quasi-Cyclic (QC) LDPC codes \cite{Tanner99qc} are structured LDPC codes defined by a base matrix $B$ with entries
$b_{i,j}\in{\mathbb N}\cup\{-1\}$. Subsequently, parity-check matrices with variable length can be obtained by
expanding the base matrix $B$ by some  factor $z\geq 1$. Within the expansion process, each entry of the base matrix is
replaced by a square $z \times z$ matrix: a $-1$ entry is replaced by the all-zero matrix, while a non-negative entry
$b_{i,j} \geq 0$ is replaced by a circulant permutation matrix corresponding to a shift by $b_{i,j}$. It is known that
non-zero entries of the base-matrix can be chosen such as to avoid unsuitable topologies in the expanded matrix (as
short cycles), which may cause degradation of the iterative decoding performance \cite{Fossorier04}.

The goal of this paper is to design LDPC codes that efficiently accommodate the hybrid IT/ML decoding. Complexity and
error correction performance of the ML decoding constitute the primary objectives. IT performance does not impact the
error correction performance of the overall scheme, but it allows for increasing throughput in the low-loss scenario.
We do not consider QC-LDPC codes for improving the IT decoding performance, but for decreasing the ML decoding
complexity. This is achieved by using a transformation of the residual system $H_e X_e = H_r X_r$ into a linear system
with a pseudo-band system matrix. This transformation exploits the quasi-cyclic structure of the parity-check matrix $H$.
Consequently, the ML decoding can be efficiently performed, and the required number of row/symbol operations scales as a
sub-quadratic power of $k$, namely $k\sqrt{k}$.

The paper is organized as follows. In Section \ref{sec:hybrid_decoding}, we briefly review the GE and ML decoding
algorithms. Band transformation and a complexity analysis of ML decoding are presented in Section \ref{sec:pseudo-band_matrix_transformation_ML_complexity}. Section \ref{sec:code_design} describes the proposed design of regular repeat-accumulate QC-LDPC codes.
Finally, Section \ref{sec:experimental_results} presents the experimental results, and Section \ref{sec:conclusion}
concludes the paper.

\section{Hybrid IT/ML decoding }
\label{sec:hybrid_decoding}

The hybrid IT/ML decoder \cite{cunche08hybride,paolini-2008a} is an advantageous combination of the IT and ML decoders,
which has the ability to cope with fluctuating channel conditions, and allows to tradeoff between complexity and
performance.

\subsection{Principles}
\label{sec:hybrid_decoding_principles}

Consider an LDPC code defined by a parity check matrix $H$, and let $X$  be a codeword transmitted over the BEC. The
subset of received symbols\footnote{Entries X of are referred to as symbols, instead of bits. Actually, in AL-FEC
applications, each symbols represents an entire packet, which is either erased or correctly received} is submitted to
the IT decoder, which may recover all or only a part of the erased symbols. If the IT decoding fails, the ML decoder is
activated, and tries to complete decoding by solving the residual system $H_e X_e=H_r X_r$, as explained in Section
\ref{sec:intro}. The system matrix $H_e$ has a number of rows equal to $m'\leq m-k$ and a number of
columns\footnote{Each symbol received or recovered by the IT decoding, removes 1 column and at least 1 row from the
system matrix} equal to $n'\leq n-k$. The above inequalities are generally tight, except when the IT decoding fails in
the error floor region (small stopping sets). This linear system can be solved by using the Gaussian elimination
method, or any other algorithm available in the literature.

\subsection{Gaussian elimination}\label{subsec:GE}
\label{sec:hybrid_decoding_resolution_algorithm}

Although many algorithms are known for solving linear systems, most of them are based on (efficient implementations of)
the Gaussian Elimination (GE) algorithm. This algorithm consists of two steps.

First, the Forward Elimination (FE) step transforms the system into an upper triangular system, which can be done as
follows. Starting from $i=0$, choose in column $i$ a non null entry, the pivot, with row-index $j\geq i$. Permute rows
$i$ and $j$, then add the row $i$ to all the rows corresponding to non-zero sub-diagonal entries of column $i$.
Simultaneously, similar operations are performed on the right-hand side of the system, {\em i.e.} the right symbol of
the $i$-th row is added to right symbols of corresponding rows.

The algorithm completes with a Backward Substitution (BS) step, which recursively recovers the last symbol of an
upper-triangular system: starting from the last column, the corresponding erased symbol is given the value of the
corresponding right-hand side symbol, and is then substituted in all the equations it is involved in.

In the remaining of the paper, this algorithm will be referred to as the \emph{``Standard Gaussian Elimination''}. Its
complexity is of order ${\cal O}(k^2)$ row/symbol operations.

\section{Pseudo-band matrix transformation and ML decoding complexity}
\label{sec:pseudo-band_matrix_transformation_ML_complexity}

It is well known that the complexity of the GE algorithm can be reduced if the system matrix is structured in some
specific way. For instance, the use of a \emph{band structure} to reduce the ML decoding complexity has been studied in
\cite{studholme_06} and \cite{Soro09ldpcband}. In this section, we show that the parity check matrix of QC-LDPC codes
features such a ``hidden'' band structure, that allows for considerably reducing the complexity of ML decoding with
standard GE.

\subsection{Transformation into a pseudo-band matrix}

Consider a base matrix $B$, of size $a\times b$, with entries from $\{-1,0,\dots,M\}$. Let $H$ be a $m\times n$ binary
matrix, obtained by expanding $B$ by some factor $z > M$; hence, $m=za$ and $n=zb$. With an appropriate row/column
permutation, the quasi-cyclic matrix $H$ can be transformed into a matrix $H'$ that exhibits a band structure.

The following algorithm performs the appropriate permutation:
\begin{enumerate}
\item[] for all $(i,j)$ in $[0, \dots, m-1]\times[0, \dots, n-1]$
  \begin{enumerate}
       \item decompose: $i=x_i z + y_i$ and $j=x_j z+ y_j$
       \item define: $i'= x_i + y_i a$ and $j'=x_j+y_j b$
       \item set: $H'[i'][j']=H[i][j]$
  \end{enumerate}
\end{enumerate}

\begin{figure}[htb]
  \begin{center}
  \includegraphics[scale=0.25]{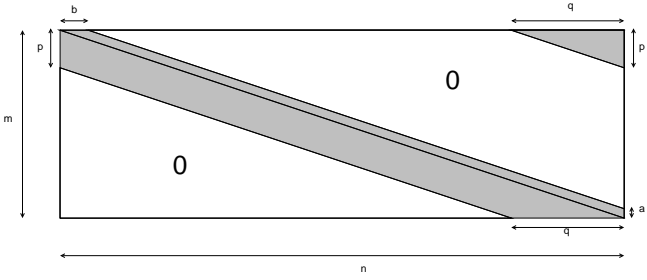}
    \end{center}\vspace{-3mm}
 \caption{$H'$, the parity check matrix after row/column permutation}
 \vspace{-5mm}
   \label{fig:H_prime_band}
\end{figure}

The resulting matrix $H'$  exhibits a pseudo-band structure, as illustrated at Figure~\ref{fig:H_prime_band}. Note
that, by convention, the $(0,0)$ position of the matrix is the bottom-right position, and the same convention will be
used for the subsequent figures. Two integers $p$ and $q$ are associated with $H'$, which represent respectively the
{\em subdiagonal} height and the width of the band. They depend on $M$, the maximum value of the non-negative entries
of $B$, and on $a$ and $b$, the dimensions of $B$. We have:
\begin{eqnarray*}
 p & = & a(M+1) \\
 q &= &  b(M+1)
\end{eqnarray*}

\begin{proof}
Consider the set of $z\times z$ circulant matrices corresponding to a right-shifted identity by $k$ positions, with $k
\in \{0, \dots, M\}$, and let $c_{\alpha,\beta}$ be the element of index $(\alpha,\beta)$ of one of these matrices.
Then $c_{\alpha,\beta}$ is potentially non-zero if and only if  $ (M \geq  \beta - \alpha \geq 0)\mbox{ or
}(\alpha-\beta\geq z-M )$. Now, $H[i][j]$ is the element with index $(y_i,y_j)$ of the $(x_i,y_j)$-th circulant matrix
composing $H$. Therefore $H[i][j]$ is potentially non-zero iff $(M \geq y_j - y_i \geq 0)$  or $(y_i-y_j \geq z-M)$.

\noindent From the first inequality, we obtain:

$$\begin{array}{rcccl}
  M & \geq &    y_j - y_i              & \geq & 0 \\
 aM & \geq & \frac{a}{b}(b y_j) - ay_i & \geq & 0 \\
\end{array}$$

\noindent In addition, we have $a \geq x_i -\frac{a}{b}x_j\geq-a$; therefore:
$$\begin{array}{rcccl}
  a(M+1) &\geq &  \frac{a}{b}(b y_j  + x_j) - ay_i - x_i &\geq & -a \\
  a(M+1) &\geq &  \frac{a}{b}j' -i'                      &\geq & -a
\end{array}$$

\noindent From the second inequality, we obtain:
$$\begin{array}{rcl}
    y_i-y_j  & \geq & z-M \\
    a y_i- \frac{a}{b} (b y_j) & \geq & az - aM \\
\end{array}$$
Again, tacking into account that $a \geq x_i -\frac{a}{b}x_j\geq-a$, we get:
$$\begin{array}{rcl}
  a y_i + x_i- \frac{a}{b} (b y_j +x_j) & \geq &az - a(M+1) \\
   i'- \frac{a}{b} j' & \geq &az - a(M+1)
\end{array}$$
Therefore $H'[i'][j']$ is potentially non-zero if and only if $( a(M+1) \geq  \frac{a}{b}j' -i' \geq -a)$  or $(i'-
\frac{a}{b} j' \geq m - a(M+1) )$, which implies that $(i', j')$ is inside the pseudo-band of $H'$.

\end{proof}
Although this  result holds for any Quasi-Cyclic code, the pseudo-band structure will be ``visible'' only if $p$ and
$q$ are significantly smaller than $m$ and $n$, respectively. This happens only if $M$ is significantly smaller than
$z$, hence, in Section \ref{sec:code_design}, we will introduce Quasi-Cyclic codes featuring an appropriate choice of
the base matrix coefficients.

\begin{figure}
 \begin{minipage}[b]{.46\linewidth}
  \includegraphics[scale=0.3]{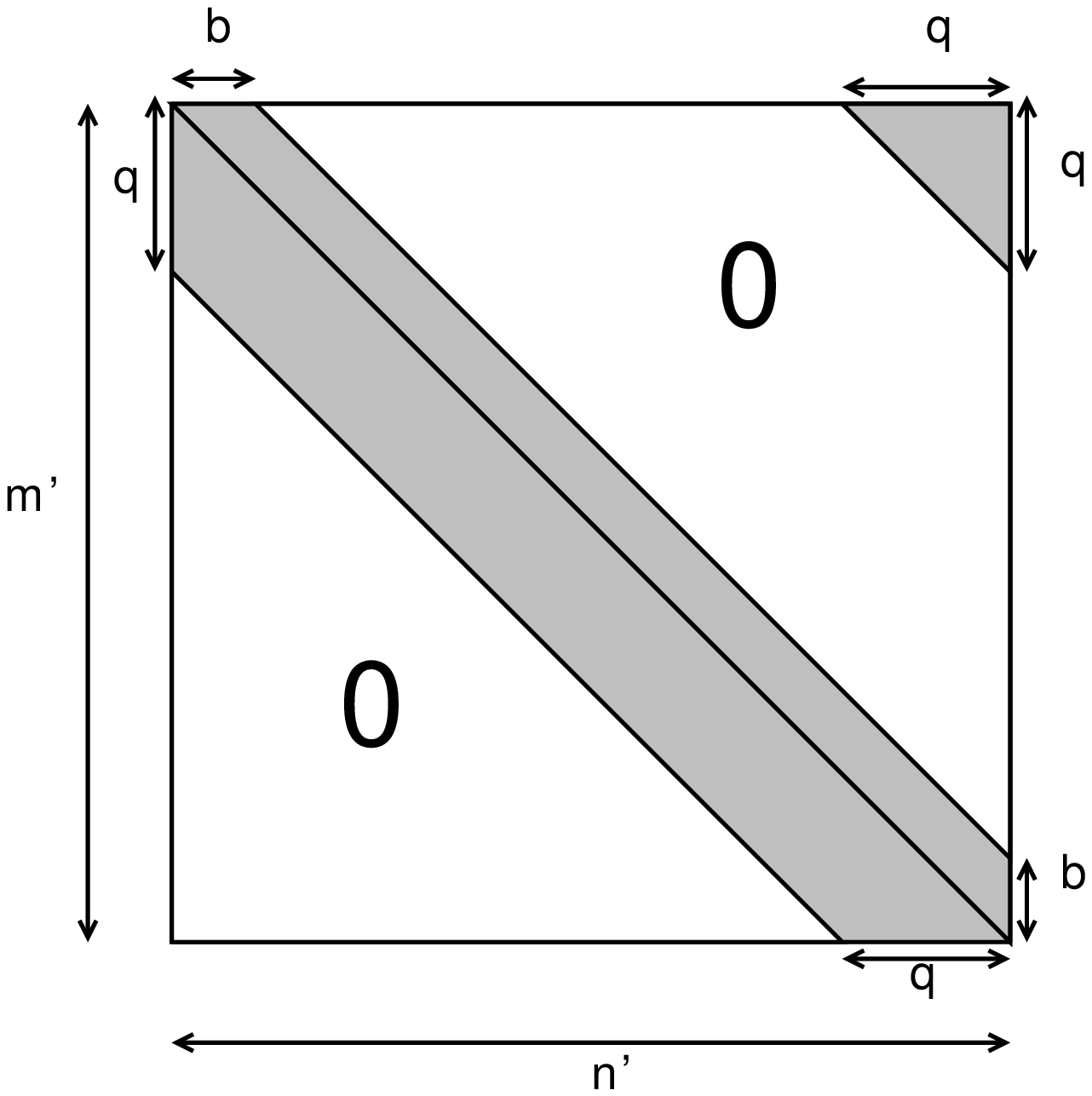}
 \caption{$H'_e$, the decoding matrix obtained from $H'$.}
   \label{fig:decoding_matrix}
 \end{minipage} \hfill
 \begin{minipage}[b]{.46\linewidth}
  \includegraphics[scale=0.3]{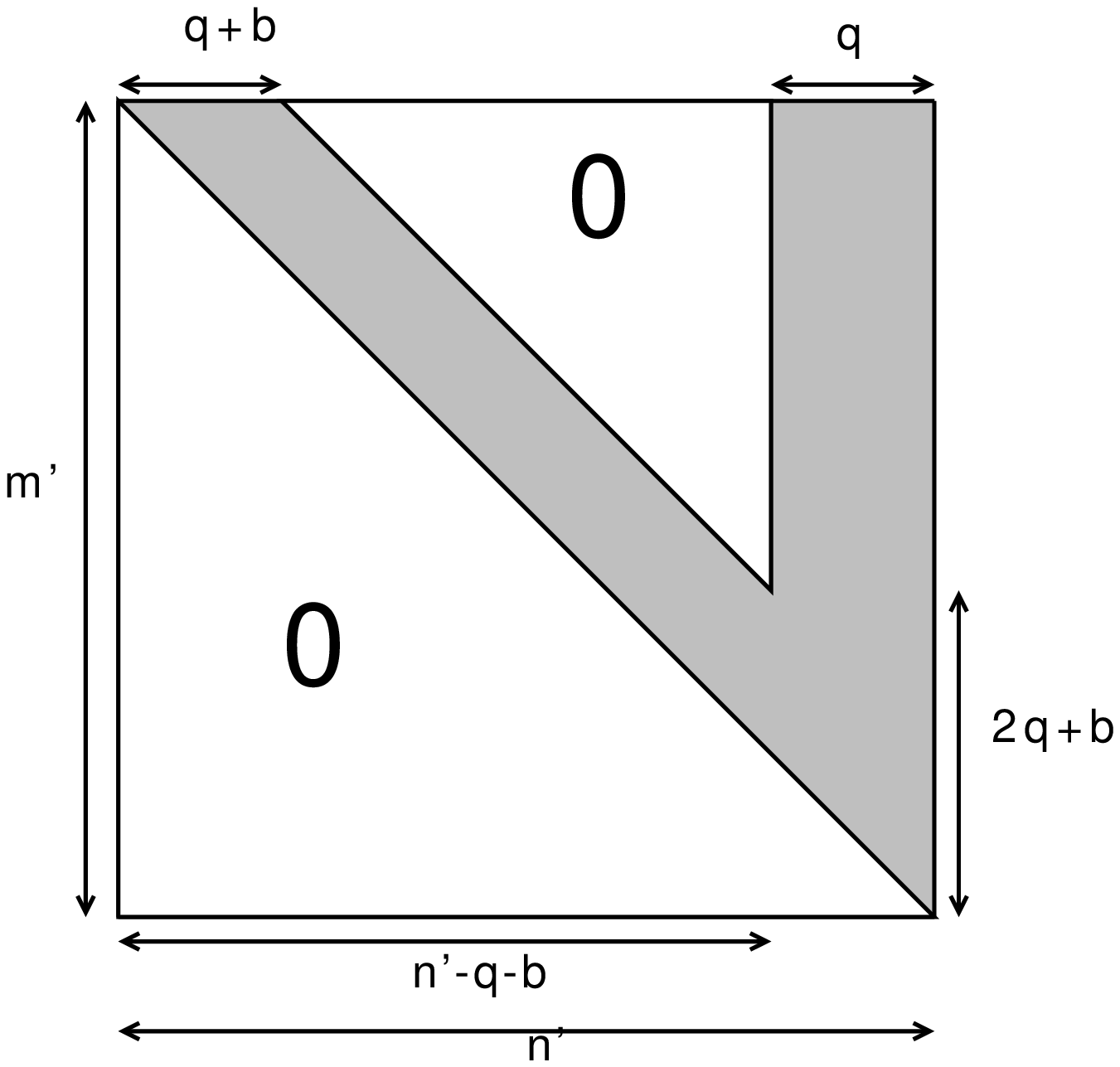}
 \caption{The decoding matrix after the Forward Elimination (FE) step.}
   \label{fig:decoding_matrix_FE}
 \end{minipage}
\end{figure}

\subsection{Complexity of Gaussian Elimination}
During the ML decoding, the linear system to be solved is represented by the decoding matrix $H'_e$, which is a
$m'\times n'$ matrix ($n'\leq m'\leq m$) composed of a subset of the rows and columns of $H'$. Consequently, $H'_e$
inherits the pseudo-band structure of $H'$, as illustrated at Figure \ref{fig:decoding_matrix}. Although the
subdiagonal height and width of the band of $H'_e$ are less than or equal to the above $p$ and $q$ parameters, for
simplicity reasons, we consider that they are both equal to $q$ (note that $q\geq p$). The same convention holds for
the supradiagonal height and width of the band, which are both considered equal to $b$. The effect of this pseudo-band
structure on the GE algorithm (Section \ref{subsec:GE}) is described below.

Thanks to the band structure of the matrix, each FE iteration ({\em i.e.} elimination of non-zero subdiagonal entries
in a column) requires only ${\cal O}(q)$ symbol operations\footnote{Remind that a symbol operation corresponds to a sum
between two rows, right-hand side term included.} per iteration. The cost of FE is therefore ${\cal O}(qn')$ symbol
operations. After the FE step, the system has a band of width $q+b$ over the diagonal (because of rows permutation),
and a column block composed of the $q$ last columns of the system (see figure \ref{fig:decoding_matrix_FE}).

Now, erased symbols are recursively recovered by the BS step, starting from the erased symbol corresponding to the last
column, back to the erased symbol corresponding to the first column. Each recovered symbol has to be substituted in the
equations it is involved in. Symbols corresponding to the last $q$ columns are each one involved in $m'$ equations,
while symbols corresponding to the first $n'-q$ columns are each one involved in $q$ equations. Therefore, the overall
cost of the BS is ${\cal O}(qm'+(n'-q)(q+b))={\cal O}(q(m'+n')+bn' -q^2-qb))$ symbol operations.

Since $q$ and $b$ are negligible with respect to $m'$ and $n'$, and $m'\approx n' \approx m$, we conclude that the
resolution of the system requires ${\cal O}((2q+b)m)$ symbol operations. Therefore the QC structure yields a complexity
gain by a factor of $m/(2q+b)$ with respect to unstructured matrices.

\section{Code design}
\label{sec:code_design}

This section focuses on the design of QC-LDPC codes, by trading-off performance and complexity constraints. Fix some
base matrix $B$ with size $a\times b$, and let $M$ be the maximum value of its non-negative entries. Using the
pseudo-band transformation of expanded matrices, it follows from the above section that the complexity of the ML
decoding scales linearly with the code dimension $k$ (or, equivalently, the expansion factor $z$). Although this is an
excellent result in terms of decoding complexity, we will see later (Section \ref{sec:experimental_results}) that for
long codes such a code design yields poor performance with both IT and ML decodings. This is explained by the fact that
the width of the pseudo-band, which depends only on $a, b$, and $M$,  becomes too thin with respect to the matrix
dimensions for large values of $z$. Such a thin band results in inappropriate graph topologies\footnote{Remind that the
pseudo-band structure is obtained by a simple row/column permutation of $H$.} for the IT decoding (more short cycles
and smaller stopping sets) and, simultaneously, it reduces the probability of $H_e$ (the ML decoding matrix) being
full-rank. In order to avoid such a situation, we propose the use of a base matrix with variable non-negative entries.
Within such a matrix, only the $-1$ entries are fixed. Equivalently, the indexes of non-negative entries are fixed, but
not their values, which may vary with the expansion factor $z$, such that to ensure that the width of the pseudo-band
is not too thin.

\paragraph*{Pseudo-band width}

In \cite{studholme_06,studholme:algorithmica}, Studholm and Blake conjectured that a matrix with a band of width
$2\sqrt{k}$, filled with $2\log{k}$ symbols per column, is full rank with probability close to that of fully random
matrices. Following this idea, we set $q=C\sqrt{k}$. This implies $M=C\sqrt{\frac{zR}{b}}$, where $R=k/n$ is the code
rate, and $C$ is a positive constant. The ML decoding with standard GE of such a code therefore requires ${\cal
O}(k\sqrt{k})$ row/symbol operations. Even if the column degree does not follow the recommendation of {\em loc. cit.},
it is chosen sufficiently large (see below) to provide excellent correction capabilities (Section
\ref{sec:experimental_results}). In addition the $C$ parameter can be adjusted to find a tradeoff between error
correction capabilities and complexity.

\paragraph*{Base matrix structure}

We use a Regular Repeat Accumulate \cite{Divsalar1998} (RRA) quasi-cyclic structure in order to benefit a linear time
encoding. The parity side of the base matrix has a double-diagonal structure, which will be referred to as {\em
staircase}. Consequently, the extended parity-check matrix inherits a staircase structure by blocks, which allows to
recursively build all the parity symbols with a linear number of symbol operations. Hybrid IT/ML decoding for Regular
Repeat Accumulate LDPC codes has been studied in \cite{Cunche.spsc08}, and more particularly the impact of the source
node degree on the decoding performance. A value of $5$ for this degree is considered as a good compromise, as it
allows excellent performance under  ML decoding, with good enough performance under IT.

\paragraph*{Base matrix entries}

The values of non-negative entries of the base-matrix are randomly chosen from $\{0,\dots,M\}$, where the maximum value
$M$ depends on the expansion factor $z$, as explained above. Such a random choice simplifies the code generation and
does not require an expensive optimization for the non-negative entry values. This is an asset when codes need to be
produced on the fly, in real time.

\paragraph*{Additional optimization}

\begin{figure}[htb]
  \begin{center}
  \includegraphics[scale=0.11]{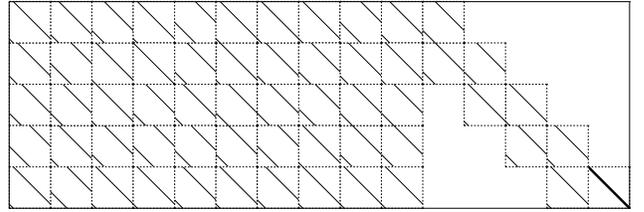}
    \end{center}\vspace{-3mm}
 \caption{Example of a QC parity check matrix (NB: the bottom right block is a staircase matrix).}
 \vspace{-5mm}
   \label{fig:qc_parity_check}
\end{figure}

If the last element of the staircase is expanded into a circulant matrix, the corresponding $z$ columns of $H$ are all
of degree one. In order to avoid the negative impact of degree one columns on the decoding performance, the last
element of the staircase is itself expanded into a staircase $z\times z$ matrix. An example of such a parity check
matrix is represented at figure \ref{fig:qc_parity_check}.

\section{Experimental results}
\label{sec:experimental_results}

We have performed experiments to assess the gains provided by the QC structure both from an erasure correction
capability and decoding complexity points of view.

\subsection{Experimental setup}

The QC-LDPC codes considered are using a base matrix having a size $5\times 15$ matrix (Figure
\ref{fig:qc_parity_check}), which is the minimum size for a rate-$2/3$ RRA matrix with a source node degree equal to
$5$.

In order to identify the influence of the QC structure and band width on the decoding performance, we consider four
code ensembles. These codes are built from the same base matrix, but using different choices for the non-negative
entries of $B$ (and also a different expansion technique for the protograph codes, see below). There are two reasons
for using a small base matrix. First, the length of the extended code is a multiple of $b$, hence, small $a$ and $b$
allow the finest grain for the length and the dimension of the extended codes. Second, the band width linearly depends
on the base matrix dimensions and $M$, which should be large enough to produce a sufficiently large range for the
random distribution of the base matrix coefficients. Therefore, for a given bandwidth, $b$ is chosen as small as
possible to maximize $M$.

\noindent The following code ensembles are considered:
\begin{mylist}
  \item \emph{band QC LDPC} codes, our proposal.
The non-negative entries $b_{i,j}$ can take any value in the range $[0, \dots, 3\sqrt{z}]$, {\em i.e.} the maximum
value $M=3 \sqrt{z}$. The factor $3$ has been chosen following a tradeoff between error correction capabilities and
complexity. These codes are QC-LDPC featuring a ``visible'' pseudo-band structure, with a width that depends on the
code dimension (Section \ref{sec:code_design}).

  \item \emph{unconstrained QC LDPC} codes.
The non-negative entries $b_{i,j}$ can take any value in the range $[0, \dots, z]$, {\em i.e.}\break $M=z$.  These
codes does not exhibit a ``visible'' pseudo-band structure (pseudo-band is too wide).

  \item \emph{constant band-width QC LDPC} codes.
The non-negative entries $b_{i,j}$ can take any value in the range $[0, \dots, M]$, where $M$ is a fixed constant,
which does not depend on the code dimension. We chose the value $M=42$ that is equal to the corresponding value for the
\emph{band QC LDPC} of dimension $k=2000$. These codes are QC-LDPC featuring a very thin pseudo-band structure, for
large values of $k$.
 \item  \emph{protograph LDPC} codes.
They are built from the same base matrix $B$, but non-negative entries are expanded into random $z\times z$ permutation
matrices, instead of circulant matrices. These codes do not have a pseudo-band structure.
\end{mylist}
For the reason presented in section~\ref{sec:code_design}, all these codes feature a $z \times z$ staircase matrix at
the bottom right. In order to avoid consideration on the loss model, the symbols are randomly permuted before the
transmission on a memoryless erasure channel. For each test the results of at least $500$ experiments is averaged.
Since we are considering code ensembles, the seed used to construct the parity check matrix is different for each
experiment.

\subsection{Erasure recovery capabilities}

\begin{figure}[!b]
  \begin{center}
  \vspace{-5mm}
\includegraphics[scale=0.9]{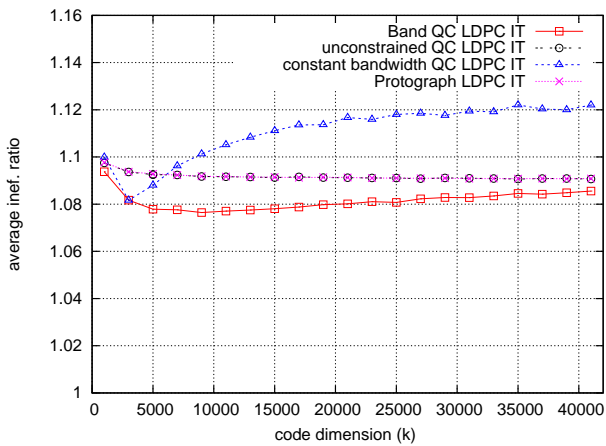}
    \end{center}\vspace{-3mm}
 \caption{Inefficiency ratio as a function of the code dimension, IT decoding ($R=2/3$).}
   \label{fig:Inefficiency_ratio_IT}
\end{figure}

\begin{figure}[!t]
  \begin{center}
\includegraphics[scale=0.9]{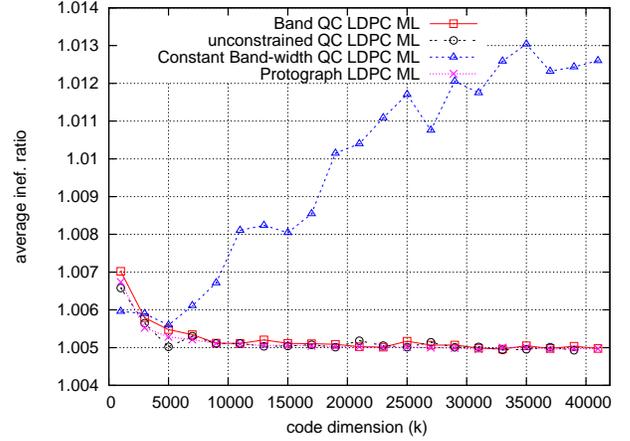}
    \end{center}\vspace{-3mm}
 \caption{Inefficiency ratio as a function of the code dimension, ML decoding ($R=2/3$).}
 \vspace{-5mm}
   \label{fig:Inefficiency_ratio_ML}
\end{figure}

The average inefficiency ratio, defined as the number of symbols required to complete decoding over the code dimension,
is presented as a function of the code dimension at figure~\ref{fig:Inefficiency_ratio_IT} for the IT decoding,  and at
figure~\ref{fig:Inefficiency_ratio_ML} for the ML decoding.

First of all, we observe that the {\em constant band-with QC LDPC} codes exhibit the worst performance, under both IT
and ML decodings. This is explained by the fact that the parity check matrix is concentrated on a pseudo-band, which is
too thin with respected to the matrix dimensions. Consequently, codes from the {\em constant band-with QC LDPC}
ensemble contain more short cycles and small stopping sets than codes from the other ensembles, which leads to a
degraded performance under the IT decoding. On the other hand, the concentration of the parity check matrix on a thin
pseudo-band decreases the probability of the ML decoding matrix being full-rank, which explains the performance under
the ML decoding.

We also observe that under the ML decoding, the average inefficiencies of \emph{Band QC LDPC}, \emph{unconstrained QC
LDPC} and \emph{protograph LDPC} are very close. Thus, even if \emph{Band QC LDPC} codes are more constrained, they are
still {\em random enough}, such as to provide ML performance close to that of unconstrained codes. This also confirms
the conjectures in \cite{studholme_06,studholme:algorithmica}, in the sense that the band width should depend on the
code dimension in order to provide ML performance close to that of unconstrained codes. Under the IT decoding, the
\emph{Band QC LDPC} codes show a slightly better inefficiency ratio than the other two code ensembles.

\begin{figure}[!htb]
  \begin{center}
\includegraphics[scale=0.9]{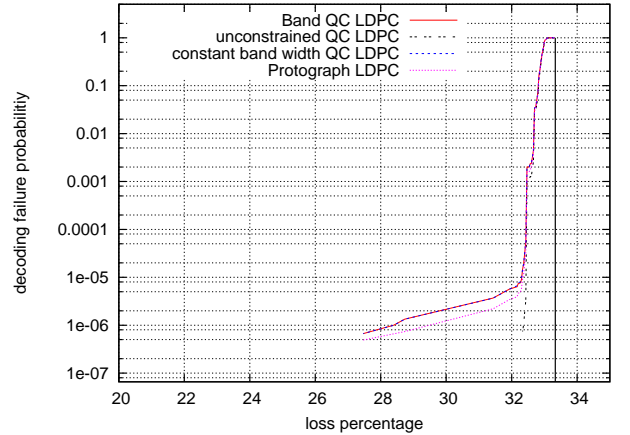}
    \end{center}\vspace{-3mm}
 \caption{Block error rate W.R.T. channel loss percentage, under ML decoding ($k=2000$, $R=2/3$).}
 \vspace{-5mm}
   \label{fig:BER_wrt_loss_percentage}
\end{figure}

Figure \ref{fig:BER_wrt_loss_percentage} shows the failure probability of the ML decoding (codeword error rate) as a
function of the loss percentage for a code dimension  $k=2000$. In the waterfall region, the different curves are
almost indiscernible and close to the theoretical limit. While no error floor is visible (down to $10^{-6}$) for
\emph{unconstrained QC LDPC} codes, the \emph{band QC LDPC}, \emph{constant band width QC LDPC} and \emph{protograph
LDPC} codes present an error floor at a failure probability of $10^{-5}$. However, this error floor is sufficiently low
for practical applications, and it is offset by a lower decoding complexity, as shown below.

\subsection{Algorithmic complexity}
\label{sec:algorithmic_complexity}

\begin{figure}[!t]
  \begin{center}
\includegraphics[scale=0.9]{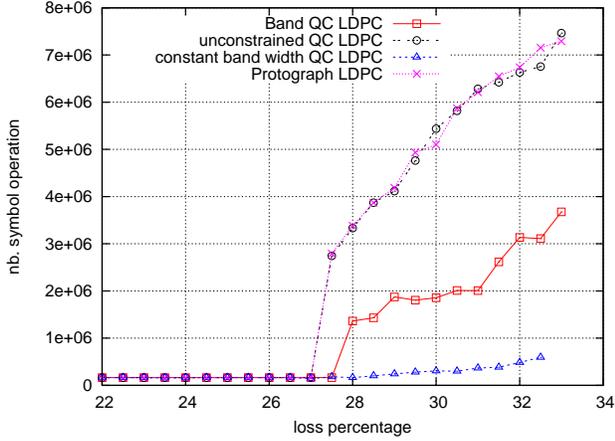}
    \end{center}\vspace{-3mm}
 \caption{Number of row/symbol operation performed during decoding W.R.T loss percentage ($k=30000$, $R=2/3$).}
 \vspace{-5mm}
   \label{fig:Nb_row_op_wrt_loss_percentage}
\end{figure}

The algorithmic complexity is evaluated by mean of number of row/symbol operations.
At figure \ref{fig:Nb_row_op_wrt_loss_percentage}, one can see that for low channel loss percentage, the number of
row/symbol operations is low (the IT decoding is sufficient). When the channel loss percentage increases, the number of
row/symbol operations increases because the ML decoding is activated more and more often. The number of operation under
IT decoding is similar for all the codes, since there parity check matrix have the same number of ones. However, once
the ML decoding is activated, the \emph{Band QC LDPC} codes  clearly outperform the \emph{protograph LDPC} and
\emph{unconstrained QC LDPC} codes. This is a direct consequence of the ``visible'' pseudo-band structure of the
decoding matrix, that allows to reduce the complexity of ML decoding. For \emph{constant band width QC LDPC} codes the
number of operations is even smaller, as their bandwidth ($q=42\times15=630$) is significantly smaller than that of the
\emph{Band QC LDPC} codes  ($q=164\times15=2460$).

\begin{figure}[!b]
  \begin{center}
\includegraphics[scale=0.9]{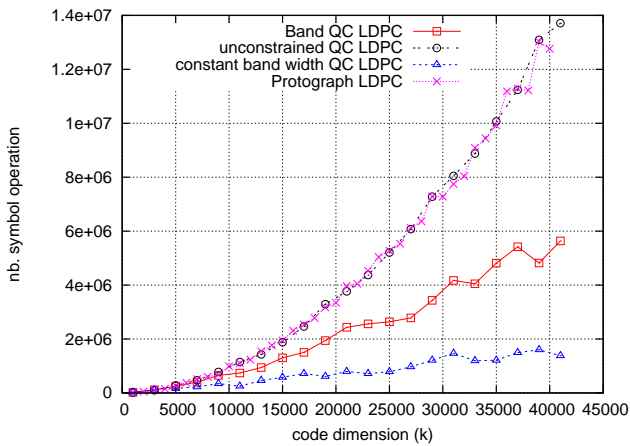}
    \end{center}\vspace{-3mm}
 \caption{Number of row/symbol operations performed during ML decoding W.R.T. the code dimension ($R=2/3$).}
   \label{fig:Nb_row_op_wrt_k}
\end{figure}

We have plotted on figure \ref{fig:Nb_row_op_wrt_k} the number of row/symbol operations performed by the ML decoding in
the worst case (minimum number of symbols received for which the ML decoding succeeds). As expected, \emph{Band QC
LDPC} and  \emph{constant band width QC LDPC} codes require fewer row/symbol operations than the other codes. The
curves of \emph{protograph LDPC} and \emph{unconstrained QC LDPC} codes are almost identical, and they do not exhibit a
specific structure that may reduce the complexity of standard GE (the pseudo-band structure of \emph{unconstrained QC
LDPC} codes is not ``visible''). This curves are  also compatible with the theoretical complexity : ${\cal O}(k)$ for
the \emph{constant band width QC LDPC} codes,  ${\cal O}(k\sqrt{k})$ for the \emph{Band QC LDPC} codes, and ${\cal
O}(k^2)$ for the \emph{protograph LDPC} and \emph{unconstrained QC LDPC}  codes.

Thus, under the ML decoding, the proposed \emph{band QC LDPC} codes perform very close to the channel capacity
(overhead of only $0.5\%$ with respect to ``the ideal code''), with tractable complexity even for large code dimension.

\section{Conclusions}
\label{sec:conclusion} In this paper we presented an analysis of the ML decoding of QC-LDPC codes over the erasure
channel. We showed that any QC matrix can be transformed into a pseudo-band form, which allows for reducing the
complexity of the ML decoding. The complexity gain depends on the ``visibility'' (width) of the pseudo-band, and the
thinner is the band, the less complex is the decoding. However, the band width has to tradeoff between performance and
complexity gain. For this end, we proposed an ensemble of QC-LDPC codes that possess excellent correction capabilities
under the ML decoding (overhead of only $0.5\%$), while decoded with a complexity of ${\cal O}(k\sqrt{k})$ in terms of
row/symbol operations. The gain in complexity increases significantly with the code dimension, which allows ML decoding
to be a realistic option for longer LDPC codes.

Additionally, the quasi-cyclic construction and the pseudo-band transformation can be generalized to any linear code
({\em i.e.} need not be low-density) in order to reduce the complexity of the ML decoding.

\bibliographystyle{plain}
\bibliography{QC_LDPC_over_BEC}

\end{document}